\documentclass[titlepage]{article}
\usepackage{longtable}
\usepackage{bm,a4wide}
\usepackage{ams}
\usepackage{amsmath,amssymb,latexsym,theorem}
\usepackage{bm}
\usepackage{rotating}
\usepackage{graphicx}
\usepackage[dvips]{color}

\begin{document}

\title{Hyperfine induced $1s2s\ ^1S_0 \rightarrow 1s^2\ ^1S_0$ M1 transition of He-like ions }

\author{Jiguang Li$^{1,2}$, Per J\"onsson$^{3}$, Gediminas Gaigalas$^{4,5}$, and Chenzhong Dong$^{1,2}$\medskip\\
$^1$ College of Physics and Electronic Engineering, Northwest Normal University, \\
Lanzhou 730070, China\\
$^2$ Joint Laboratory of Atomic and Molecular Physics, NWNU \& IMPCAS, \\
Lanzhou 730070, China\\
$^3$ Nature, Environment, Society, Malm\"o University, \\
Malm\"o S-20506, Sweden\\
$^4$ Department of Physics, Vilnius Pedagogical University, \\
Studentu 39, Vilnius LT-08106, Lithuania\\
$^5$ Institute of Theoretical Physics and Astronomy, \\
A. Gostauto 12, Vilnius LT-01108, Lithuania}

\date{}

\maketitle

\begin{abstract}
Hyperfine induced $1s2s\ ^1S_0 \rightarrow 1s^2\ ^1S_0$ M1
transition probabilities of He-like ions have been calculated from
relativistic configuration interaction wavefunctions including
the frequency independent Breit interaction and QED effects.
Present results for {$^{151}$}Eu and {$^{155}$}Gd are in good agreement with previous
calculations [Phys. Rev. A {\bf 63}, 054105 (2001)].
Electronic data are given in terms of a general scaling law in $Z$ that, given
isotopic nuclear spin and magnetic moment, allows hyperfine induced
decay rates to be estimated for any isotope. The results should
be helpful for future experimental investigations on QED and parity
non-conservation effects.

\vspace{1cm}

\end{abstract}

\pagestyle{empty}

\thispagestyle{empty}


The hyperfine structure of atomic levels is caused by the
interaction between the electrons and the non-central
electromagnetic multipoles of the nucleus. The interaction, although
weak, not only shifts and splits the individual $J$ levels, but also
mixes wavefunctions with different $J$ quantum numbers
\cite{Armstrong}. The mixing may open new decay channels and induce
radiative transitions such as $J = 0 \rightarrow J'=0$. The study of
hyperfine induced transitions (HIT) is attractive in view of getting
a more subtle understanding of electron correlation and relativistic
effects \cite{Aboussaid, Cheng, Liu}, testing the standard model
\cite{Bennett}, obtaining nuclear properties \cite{Toleikis,
Labzowsky, Indelicato}, developing ultra-precise atomic clocks
\cite{Becker, Takamoto, Rosenband, Porsev}, analyzing stellar
spectra \cite{Andersson}, and determining isotopic abundance ratios and
electron densities of plasma \cite{Brage,Brage2}.

He-like systems are interesting due to their relative simplicity, and important effects such as electron
correlation \cite{Fischer}, quantum electrodynamics (QED)
\cite{Marrus, Johnson, Indelicato2} and parity non-conservation (PNC) phenomena \cite{Gorshkov,
Maul, Dunford, Labzowsky2} can be investigated in detail. Gorshkov
and Labzovski$\rm \check{i}$ \cite{Gorshkov} and Labzowsky {\it et
al.} \cite{Labzowsky2} have proposed that the mixed hyperfine- and
weak-quenching can be used to test parity-violation effects. The one
photon transition $1s2s~^1S_0 \rightarrow 1s^2~^1S_0$ of He-like
ions is considered a good candidate for these tests and experiments will be carried
out at GSI \cite{Bondarevskaya}. Therefore,
accurate hyperfine induced $1s2s~^1S_0 \rightarrow 1s^2~^1S_0$ M1
transition probabilities of He-like ions are important. Relevant data
are however still insufficient
and in response to this we have performed systematic calculations along the He-like
iso-electronic sequence using GRASP2K \cite{grasp2K} based on the multi-configuration Dirac-Hartree-Fock
method and the HFST \cite{HFST} package. From computed hyperfine induced transitions for 14 ions and
corresponding electronic quantities we have derived a scaling formula in $Z$ that allows the induced
transition rate to be computed for any isotope.

When the hyperfine interaction is included the wavefunctions of
the combined electronic and nuclear system can be represented as
\begin{eqnarray}
|\Gamma F M_F \rangle = \sum_{i} c_i |\gamma_i I J F M_F \rangle.
\end{eqnarray}
The zero-order functions $|\gamma I J F M_F \rangle$ in the expansion
are coupled products of
electronic $|\gamma J M_J\rangle$ and
nuclear $|I M_I \rangle $ wavefunctions. The $1s^2~^1S_0$ ground state is well
represented by a single term. For $1s2s~^1S_0$ only
the interaction with $1s2s~^3S_1$ is important (see Figure 1) and
the wavefunction can be
approximated by the expansion
\begin{equation}
|``1s2s~^1S_0\, I F "  \rangle = c_0 | 1s2s~^1S_0\, I F \rangle + c_1 |1s2s~^3S_1\,I F \rangle,
\end{equation}
where $I$ is the nuclear spin and $F(=I)$ the total angular momentum quantum number. Magnetic quantum numbers are suppressed for brevity.
The use of quotation marks in the left-hand wavefunction emphasizes the fact that
the notation is just a label indicating the dominant character of the eigenvector.
The mixing coefficient $c_1$ is obtained in first order perturbation theory
as the ratio between the hyperfine matrix element and the unperturbed energy differences
\begin{equation}
c_1 = \frac{\langle 1s2s~^3S_1\, I F | H_{hfs} |1s2s~^1S_0\, I F \rangle}{E(^1S_0) - E(^3S_1)}.
\end{equation}
The one-photon $1s2s~^1S_0 \rightarrow 1s^2~^1S_0$ M1 transition becomes allowed via the mixing of
$1s2s~^3S_1$ and the decay rate in s$^{-1}$ is given by
\begin{eqnarray}
A_{HIT}(1s2s~^1S_0 \rightarrow 1s^2~^1S_0) =
\frac{2.69735 \times 10^{13}\,c_1^2\,S_{M1}}{3 \lambda^3},
\end{eqnarray}
where
\begin{equation}
S_{M1} = \left| \langle 1s^2~^1S_0 \| {\bf M}^{(1)}  \| 1s2s~^3S_1 \rangle \right|^2
\end{equation}
is the line strength computed from unperturbed wavefunctions.
$\lambda$ is the wavelength in {\AA} for the transition. The reader
is referred to \cite{Brage, Akhiezer, Edmonds} for details of the derivation.

The electronic wavefunctions were computed using the
GRASP2K program package \cite{grasp2K}. Here the wavefunction for a state labeled $\gamma J$ is approximated by an expansion
over $jj$-coupled configuration state functions (CSFs)
\begin{equation}
|\gamma J \rangle = \sum_i d_i \Phi_i.
\end{equation}
In the multi-configuration self-consistent field (SCF) procedure both the radial parts of the orbitals and the expansion coefficients
are optimized to self-consistency. In the present work a Dirac-Coulomb Hamiltonian was used with the nucleus described
by an extended Fermi charge distribution \cite{Parpia}. The multi-configuration SCF calculations
were followed by relativistic CI calculations including the frequency independent Breit interaction
and leading QED effects.


For the low charged ions the main uncertainties in the calculation
come from electron correlation effects. To build a reasonable
correlation model and control the accuracy we performed tentative
calculations of transition energies and the $\langle 1s2s~^3S_1\, I F | H_{hfs} |1s2s~^1S_0\, I F \rangle$ off-diagonal hyperfine matrix element
in $^{12}$C. In the calculations the wavefunctions for $1s^2~^1S_0$, $1s2s~^1S_0$, $1s2s~^3S_1$ were
determined simultaneously in extended optimal level (EOL)
calculations \cite{Dyall}. All CSFs that could be built from an
active set of orbitals were included in the expansion. The orbital
set was systematically increased by adding layers of new orbitals.
The largest active set included relativistic orbitals with principal
quantum number $n \le 7$. Due to stability problems in the
relativistic SCF procedure only the outermost layers of orbitals
could be optimized each time. The frequency independent Breit interaction
and leading QED effects were added in subsequent relativistic CI calculations.
The results for $^{12}$C are shown in Table 1. The first
column in this table represent the active set of orbitals involved
in each step of the calculation. As can be seen from the table, five
energy-optimized layers of orbitals are needed to converge the
off-diagonal hyperfine interaction matrix element between $1s2s~^3S_1$ and $1s2s~^1S_0$ at a sub per mill level.

Based on the above analysis, we performed calculations for other He-like ions using an active orbital
set with $n \le 5$. The Breit interaction and main QED corrections were included. Values of nuclear
magnetic dipole moments for the different isotopes were adopted
from the compilation by Stone \cite{Stone}. In Table 2 we display the
off-diagonal hyperfine matrix elements and corresponding mixing coefficients $c_1$.
The given values are not corrected for the
the anomalous magnetic moment. For the
mixing coefficient of $^{151}$Eu the magnitude is agreement with the
value of \cite{Labzowsky2}, 
but with a difference in sign due to different definitions
of the phase factor in the hyperfine interaction matrix element. The difference in sign
does not influence the final hyperfine induced transition probability.
Wavelengths $\lambda$ and line strengths $S_{M1}$ for the $1s2s\ ^3S_1 - 1s^2\ ^1S_0 $
M1 transition needed for the quenching rate are taken from accurate relativistic CI calculations by Johnson {\em et al.} \cite{Johnson}.
The values are presented in Table 3 for the convenience of the reader.
The hyperfine induced $1s2s ^1S_0 \rightarrow 1s^2\ ^1S_0$ M1 transition
rate and corresponding wavelengths of He-like ions are given Table 4. Previous theoretical results of wavelength \cite{Plante} are compared with
present calculations in this table. The agreement between our value for $^{151}$Eu and previous theoretical values by
Labzowsky {\em et al.} \cite{Labzowsky2} is very good.

To predict the transition rate for any isotope in the iso-electronic sequence we follow Brage {\it et al.} \cite{Brage}
and factorize the hyperfine induced transition rate into nuclear and electronic parts
\begin{equation}
A_{HIT}(1s2s~^1S_0 \rightarrow 1s^2~^1S_0) = \mu_I^2 (1+ 1/I) A^{el}(1s2s~^1S_0 \rightarrow 1s^2~^1S_0).
\end{equation}
The electronic part $A^{el}$ has a smooth behavior along the iso-electronic
sequence making interpolation possible. From the data in Table 4 we obtain a fit of the form
\begin{eqnarray}
A^{el} = 1.9728 \times 10^{-19}\,  Z^{14.065}
\end{eqnarray}
where $Z$ is the atomic number. The fit is shown in Figure 2. In
order to show clearly the trend of $A^{el}$ with Z, log(A) was
plotted in this picture. Using the fitting formula we estimate the
probability of $^{155}$Gd with nuclear spin $I=3/2$ and nuclear
dipole moment $\mu=-0.2591 \mu_N$ to 5.60 $\times 10^{5}$ s$^{-1}$.
This value is in good agreement with the theoretical value 5.8
$\times 10^{5}$ s$^{-1}$ given in \cite{Labzowsky2}. The fitting
formula (8) is expected to provide accurate values, but further
examination through experiment and theory is still needed.

To sum up we have calculated hyperfine induced $1s2s\ ^1S_0
\rightarrow 1s^2\ ^1S_0$ M1 transition probabilities of He-like ions
using GRASP2K \cite{grasp2K} based on multi-configuration Dirac-Fock
method and the HFST \cite{HFST} package. Electron
correlation effects were included in a systematic way. The Breit interaction and QED
effects were included in subsequent relativistic CI calculations. A scaling law in $Z$ was
derived for the electronic quantities. The scaling law allows hyperfine induced
transition probabilities to be estimated for any isotope.

This work has been supported by the National Nature Science
Foundation of China (Grant No. 10434100, 10774122), the specialized
Research Fund for the Doctoral Program of Higher Education of China
(Grant No. 20070736001) and the Foundation of Northwest Normal
University (NWNU-KJCXGC-03-21). Financial support by the Swedish
Research Council is gratefully acknowledged.

\clearpage

\begin{figure}
\centering
\includegraphics[scale=1]{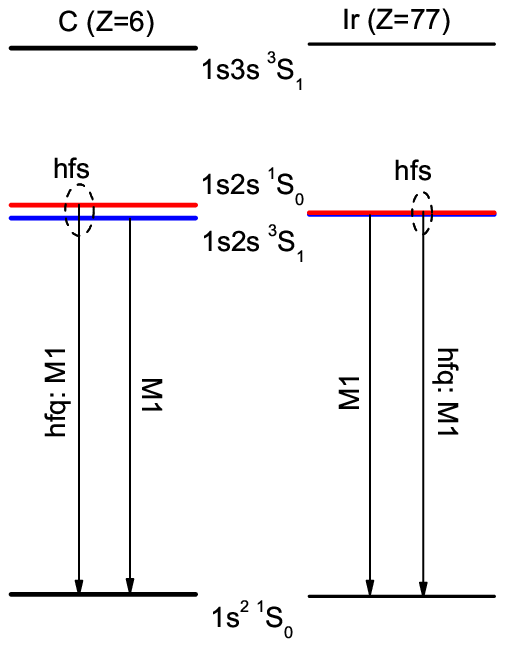}
\caption{Level structure of the $1s^2$, $1s2s$ and $1s3s$
configurations in He-like C and Ir.}
\end{figure}

\begin{table}
\caption{Transition energies in cm$^{-1}$ and off-diagonal
hyperfine interaction matrix element in a.u. for ${^{12}}$C from calculations with increasing active sets.
$\Delta E_1 = E(1s2s\ ^3S_1)-E(1s^2\ ^1S_0)$,
$\Delta E_2 = E(1s2s\ ^1S_0)-E(1s2s\ ^1S_0)$,
$\Delta E_3 = E(1s2s\ ^1S_0)-E(1s2s\ ^3S_1)$.}
\begin{tabular}{lcccc} \hline
active set      & $\Delta E_1$ & $\Delta E_2$ &$\Delta E_3$& Matrix Element  \\ \hline
2s1p            &  2407416   &  2451264   & 43847  & 4.3613[-6]     \\
3s2p1d          &  2410850   &  2455165   & 44315  & 4.5773[-6]     \\
4s3p2d1f        &  2411282   &  2455181   & 43899  & 4.5641[-6]     \\
5s4p3d2f1g      &  2411308   &  2455148   & 43840  & 4.5576[-6]    \\
6s5p4d3f2g1h    &  2411422   &  2455224   & 43802  & 4.5576[-6]    \\
7s6p5d4f3g2h1i  &  2411475   &  2455265   & 43791  & 4.5575[-6]    \\
\hline
QED corrections  &  2411439   &  2455237   & 43798  & 4.5572[-6]    \\
\hline
Experiment\cite{NIST}&  2411271   &  2455026   & 43755  &           \\
\hline
\end{tabular}
\end{table}

\begin{table}
\caption{Off-diagonal hyperfine interaction matrix elements (a.u.) and hyperfine mixing coefficients $c_1$ of He-like ions.
Nuclear magnetic dipole moment $\mu_I$ are from \cite{Stone}.}
\begin{tabular}{lcccrrr} \hline
&&&&&\multicolumn{2}{c}{Mixing Coefficient} \\
\cline{6-7} Isotope    &  $Z$  & $I$      & $\mu_I$  & \multicolumn{1}{c}{Matrix Element} & This work & Ref.\cite{Labzowsky2}        \\
\hline
$^{12}$C    & 6   & 1/2 & 0.7042      &  4.5576[-6]  &  2.2817[-5]  &       \\
$^{19}$F    & 9   & 1/2 & 2.628868    &  5.7260[-5]  &  1.6993[-4]  &       \\
$^{28}$Si   & 14  & 1/2 & -0.55529    & -4.5667[-5]  & -8.1505[-5]  &       \\
$^{47}$Ti   & 22  & 5/2 & -0.78848    & -1.7489[-4]  & -1.8463[-4]  &       \\
$^{57}$Fe   & 26  & 1/2 & 2.56277     &  4.8981[-5]  &  4.2385[-5]  &       \\
$^{70}$Ga   & 31  & 3/2 & 2.6289      &  1.7944[-3]  &  1.2473[-3]  &       \\
$^{85}$Rb   & 37  & 5/2 & 1.35298     &  1.5213[-3]  &  8.4420[-4]  &       \\
$^{97}$Mo   & 42  & 5/2 & -0.9335     & -1.5826[-3]  & -7.3716[-4]  &       \\
$^{103}$Rh  & 45  & 1/2 & -0.884      & -2.7542[-3]  & -1.1627[-3]  &       \\
$^{117}$Sn  & 50  & 1/2 & -1.00104    & -4.4439[-3]  & -1.6030[-3]  &       \\
$^{131}$Xe  & 54  & 3/2 & 0.6908      &  2.9782[-3]  &  9.5300[-4]  &       \\
$^{151}$Eu  & 63  & 5/2 & 3.4717      &  2.3846[-2]  &  5.8619[-3]  & -5.83[-3]\\
$^{175}$Lu  & 71  & 7/2 & 2.2323      &  2.3187[-2]  &  4.5372[-3]  &       \\
$^{193}$Ir  & 77  & 3/2 & 0.1637      &  2.6895[-3]  &  4.4400[-4]  &       \\
\hline
\end{tabular} \\
\end{table}

\begin{table}
\caption{Wavelengths $\lambda$ in {\AA}, rates $A_{M1}$ in s$^{-1}$ and line strengths $S_{M1}$ in a.u. for the
$1s2s~^3S_1 - 1s^2~^1S_0$ M1 transition of He-like ions. From Johnson {\em et al.} \cite{Johnson}.}
\begin{tabular}{lcccc} \hline
Ion & $Z$ & \multicolumn{1}{c}{$\lambda$} & \multicolumn{1}{c}{$A_{M1}$} &
\multicolumn{1}{c}{$S_{M1}$} \\
\hline
C   & 6  & 41.470   & 4.860[1] & 3.8550[-7] \\
F   & 9  & 17.152   & 3.621[3] & 2.0323[-6] \\
Si  & 14 & 6.7402   & 3.598[5] & 1.2253[-5] \\
Ti  & 22 & 2.6368   & 3.750[7] & 7.6467[-5] \\
Fe  & 26 & 1.8682   & 2.075[8] & 1.5046[-4] \\
Ga  & 31 & 1.3002   & 1.257[9] & 3.0719[-4] \\
Rb  & 37 & 0.90272  & 7.714[9] & 6.3110[-4] \\
Mo  & 42 & 0.69459  & 2.842[10]& 1.0592[-3] \\
Rh  & 45 & 0.60198  & 5.792[10]& 1.4054[-3] \\
Sn  & 50 & 0.48342  & 1.726[11]& 2.1681[-3] \\
Xe  & 54 & 0.41151  & 3.846[11]& 2.9807[-3] \\
Eu  & 63 & 0.29715  & 1.943[12]& 5.6695[-3] \\
Lu  & 71 & 0.22993  & 6.950[12]& 9.3969[-3] \\
Ir  & 77 & 0.19266  & 1.672[13]& 1.3300[-2] \\
\hline
\end{tabular} \\
\end{table}

\begin{table}
\caption{Hyperfine induced rates $A_{HIT}$ in s$^{-1}$ and corresponding wavelength $\lambda$ in {\AA}
for the $1s2s ^1S_0 \rightarrow 1s^2\ ^1S_0$ transition in He-like ions.}
\begin{tabular}{lcccccccc} \hline
&&&&\multicolumn{2}{c}{$\lambda$(\AA)}&&\multicolumn{2}{c}{$A_{HIT}$}\\
\cline{5-6}\cline{8-9}Isotope     & $Z$  & $I$ & $\rm \mu_I$& This work & Ref.\cite{Plante} && This work & Ref.\cite{Labzowsky2}        \\
\hline
$^{12}$C    & 6   & 1/2 & 0.7042     & 40.7306 & 40.7304 && 2.6704[-8]  &   \\
$^{19}$F    & 9   & 1/2 & 2.628868   & 16.9361 & 16.9404 && 1.0862[-4]  &   \\
$^{28}$Si   & 14  & 1/2 & -0.55529   & 6.6835  & 6.6848  && 2.4515[-3]  &   \\
$^{47}$Ti   & 22  & 5/2 & -0.78848   & 2.6214  & 2.6225  && 1.3010[0]  &   \\
$^{57}$Fe   & 26  & 1/2 & 2.56277    & 1.8584  & 1.8594  && 3.7865[-1]  &   \\
$^{70}$Ga   & 31  & 3/2 & 2.6289     & 1.2940  &         && 1.9830[3] &   \\
$^{85}$Rb   & 37  & 5/2 & 1.35298    & 0.8987  &         && 5.5705[3] &   \\
$^{97}$Mo   & 42  & 5/2 & -0.9335    & 0.6916  & 0.6923  && 1.5643[4] &   \\
$^{103}$Rh  & 45  & 1/2 & -0.884     & 0.5994  &         && 7.9312[4] &   \\
$^{117}$Sn  & 50  & 1/2 & -1.00104   & 0.4814  & 0.4820  && 4.4904[5] &   \\
$^{131}$Xe  & 54  & 3/2 & 0.6908     & 0.4098  & 0.4104  && 3.5374[5] &   \\
$^{151}$Eu  & 63  & 5/2 & 3.4717     & 0.2958  &         && 6.7643[7] & 6.8[7]  \\
$^{175}$Lu  & 71  & 7/2 & 2.2323     & 0.2289  &         && 1.4508[8] &   \\
$^{193}$Ir  & 77  & 3/2 & 0.1637     & 0.1917  &         && 3.3463[6] &   \\
\hline
\end{tabular} \\
\end{table}

\begin{figure}
\centering
\includegraphics[scale=1]{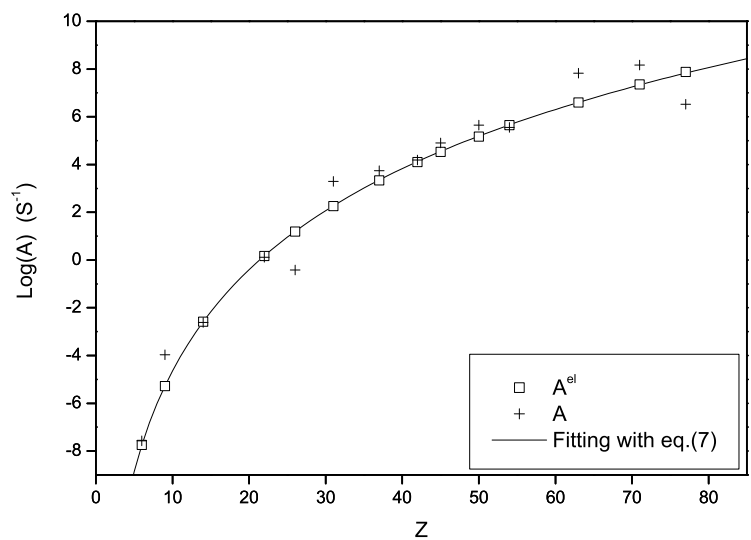}
\caption{Logarithm of hyperfine induced rates $A_{HIT}$ for the
$1s2s ^1S_0 \rightarrow 1s^2\ ^1S_0$ transition of He-like ions
together with fit to the corresponding electronic quantity
$A^{el}$.}
\end{figure}

\end{document}